\documentclass[12pt]{article}

\usepackage{amsmath,amssymb}
\usepackage{graphicx}
\usepackage{caption}
\usepackage{color}
\usepackage{dcolumn}
\usepackage{bm}
\usepackage[numbers,super,comma,sort&compress]{natbib}

\definecolor{background-color}{gray}{0.98}

\title{Machine Learning for Atomic Forces in a Crystalline Solid: Transferability to Various Temperatures}
\author{Teppei Suzuki\thanks{International Center for Materials Nanoarchitectonics, National Institute for Materials Science, 1-1, Namiki, Tsukuba, Ibaraki, 305-0044, Japan, 
SUZUKI.Teppei@nims.go.jp}, 
Ryo Tamura\thanks{International Center for Materials Nanoarchitectonics, National Institute for Materials Science, 1-1, Namiki, Tsukuba, Ibaraki, 305-0044, Japan, 
Center for Materials Research by Information Integration,
National Institute for Materials Science, 1-2-1 Sengen, Tsukuba, Ibaraki 305-0047, Japan, 
TAMURA.Ryo@nims.go.jp}, 
Tsuyoshi Miyazaki\thanks{International Center for Materials Nanoarchitectonics, National Institute for Materials Science, 1-1, Namiki, Tsukuba, Ibaraki, 305-0044, Japan, 
Center for Materials Research by Information Integration,
National Institute for Materials Science, 1-2-1 Sengen, Tsukuba, Ibaraki 305-0047, Japan}}

\begin{document}

\maketitle

\begin{abstract}
Recently, machine learning has emerged as an alternative, powerful approach for predicting quantum-mechanical properties of molecules and solids. 
Here, using kernel ridge regression and atomic fingerprints representing local environments of atoms, we trained a machine-learning model on a crystalline silicon system in order to directly predict the atomic forces at a wide range of temperatures. 
Our idea is to construct a machine-learning model using a quantum-mechanical data set taken from canonical-ensemble simulations at a higher temperature, or an upper bound of the temperature range. 
With our model, the force prediction errors were about 2\% or smaller with respect to the corresponding force ranges, in the temperature region between 300 and 1650 K. 
We also verified the applicability to a larger system, ensuring the transferability with respect to system size. 
\end{abstract}

\clearpage


  \makeatletter
  \renewcommand\@biblabel[1]{#1.}
  \makeatother

\bibliographystyle{apsrev}

\renewcommand{\baselinestretch}{1.5}
\normalsize

\clearpage


\section*{\sffamily \Large INTRODUCTION}

Understanding thermodynamic properties of materials, especially nanomaterials such as nanowires, is essential for designing and manufacturing new devices.\cite{Volz-1999}
One of the most accurate and reliable methods to understand such materials at the atomistic level is the use of molecular dynamics (MD) simulations based on density functional theory (DFT). 
Unfortunately, the length and time scales needed for the prediction of thermodynamic and kinetic properties using {\it ab initio} MD methods are often beyond the reach of present-day computer power. 
Examples include the evaluation of dynamical activation energy\cite{Boisvert-1996} or thermal conductivity,\cite{Plathe-1997} where a number of simulations on large, realistic models at different temperatures on the time scale of hundreds of picoseconds are required, which makes {\it ab initio} MD studies of these properties practically prohibitive. 
Many classical force-field simulations have therefore been applied to larger systems for longer time scales\cite{Volz-1999,Plathe-1997,Wang-2009,Ju-2012}; 
however, a major drawback is that empirical potentials often suffer from the transferability to chemically complex environments and to higher temperatures.\cite{Li-2015}

Recently, machine-learning (ML) approaches have been applied to predicting a variety of properties of molecules and solids: atomization energies,\cite{Hansen-2013,Rupp-2015} nuclear chemical shifts,\cite{Rupp-2015a} 
interatomic potentials,\cite{Behler-2007,Bartok-2010,Behler-2011,Behler-2011a,Seko-2015,Artritha-2016} and force constants.\cite{Tadano-2014,Zhou-2014,Tadano-2015} 
ML methods employed in these efforts include artificial neural networks,\cite{Behler-2007,Behler-2011,Behler-2011a} Gaussian process regression,\cite{Bartok-2010,Li-2015} 
compressive sensing,\cite{Zhou-2014,Tadano-2015}
and kernel ridge regression (KRR).\cite{Hansen-2013,Rupp-2015a,Rupp-2015} 
Interestingly, ML approaches have been successful in {\it direct} predictions of atomic forces for one- and two-component solid-state systems,\cite{Botu-2015a,Botu-2015b,Li-2015} with a small fraction of the computational cost needed for quantum mechanical (QM) evaluation. 
Moreover, their predictive power is often on par with that of DFT\cite{Botu-2015a,Botu-2015b,Li-2015}; hence, recent progress in data-driven, ML force fields\cite{Behler-2011a,Li-2015,Botu-2015a,Botu-2015b,Caccin-2015,Shaughnessy-2016} is quite encouraging. 
With this perspective, constructing ML force fields that can be transferable across a broad range of temperatures is an essential ingredient in the development of fast and reliable ML-based MD methods. 
To our knowledge, however, the information about assessing the quality of ML force fields in terms of temperature has been elusive; yet, it is not {\it a priori} obvious how well a trained ML model can predict atomic forces at different temperatures.

To address the above topic, herein we provide a simple, intuitive prescription for generating a versatile data set for training a robust ML force field that can be applicable to a range of temperatures. 
Our idea is based on two premises: (i) MD trajectories of a solid system will revisit similar regions in the phase space; and (ii) with the help of a proper atomic representation and a nonlinear ML technique, atomic forces in a crystalline solid at {\it different} temperatures can be accurately predicted, using a data set obtained from the canonical ensemble at a {\it much higher} temperature. 
In this paper, we argue that our ML force field, once carefully trained by a QM data set chosen from a high-temperature simulation in the canonical ensemble, can predict atomic forces in a crystalline solid across a range of temperatures.

This paper is organized as follows. In the Methodology section, first we briefly describe the KRR method, followed by a cross-validation scheme, which assesses the quality of our ML model and determines the optimal values for hyperparameters. 
Second, we give a brief overview of a descriptor that can simply and efficiently represent local atomic environments, called {\it atomic fingerprint}, which has been recently introduced by Botu and Ramprasad.\cite{Botu-2015a,Botu-2015b}
Third, we present the computational details for generating QM data sets and provide a definition for evaluating the force error. 
In the Results and Discussion section, after verifying a merit of using a training data set taken from the MD trajectory at the upper bound of the temperature range, we investigate how the training data set size and the fingerprint complexity affect the prediction error. 
Then, we show how accurately our ML model on crystalline silicon (trained only by a QM data set at 1650 K) can predict the atomic forces at different temperatures for the same system size and for a larger one. 
Finally, we summarize the conclusions.


\section*{\sffamily \Large METHODOLOGY}

\section*{\sffamily \Large Kernel ridge regression}

KRR is a kernelized version of linear ridge regression, where the nonlinearity is embedded by mapping the data into a high-dimensional Hilbert space, called {\it feature space}.\cite{Rupp-2015}
The key idea of kernel-based ML, known as {\it kernel trick}, is to implicitly express the inner product in feature space via a chosen kernel without explicitly carrying out the transformation to feature space.
For an introduction to KRR in the context of predicting QM properties, see a tutorial review by Rupp.\cite{Rupp-2015} 
KRR has been successfully applied to materials and chemical sciences.\cite{Pilania-2013,Hansen-2013,Botu-2015a,Botu-2015b,Rupp-2015a} 
With KRR, a prediction $F^* (\mathbf{X})$ is given by\cite{Rupp-2015}
\begin{eqnarray}
F^* (\mathbf{X}) = \mathbf{k}^{\rm T} (\mathbf{K} + \lambda \mathbf{I}_N)^{-1} \mathbf{F},
\end{eqnarray}
with
\begin{eqnarray}
\mathbf{F} &=& \begin{pmatrix}  F (\mathbf{X}_1) & \cdots & F (\mathbf{X}_N) \end{pmatrix}^{\rm T}, \\
\mathbf{k} &=& \begin{pmatrix} k (\mathbf{X}_1, \mathbf{X}) & \cdots & k (\mathbf{X}_N, \mathbf{X}) \end{pmatrix}^{\rm T}, \\
\mathbf{K} &=& \begin{pmatrix} 
k (\mathbf{X}_1, \mathbf{X}_{1}) & \cdots & k (\mathbf{X}_1, \mathbf{X}_{N}) \\
\vdots & \ddots & \vdots \\
k (\mathbf{X}_N, \mathbf{X}_{1}) & \cdots & k (\mathbf{X}_N, \mathbf{X}_{N})
\end{pmatrix},
\end{eqnarray}
where $\lambda$ is a hyperparameter that determines the strength of regularization, $N$ is the number of training data $\{ \mathbf{X}_n, F (\mathbf{X}_n) \} \, (n=1, ..., N)$, $\mathbf{I}_N$ denotes the $N \times N$ identity matrix, and $k (\mathbf{X}_n, \mathbf{X}_m)$ \ is the kernel. 
Note that, with KRR, the computational cost of interpolation scales linearly with the number of training data,\cite{Botu-2015b} once the training phase is properly conducted. 
While different kernel functions can be used,\cite{Hansen-2013,Rupp-2015} in this work, we used one of the most popular kernels, namely the Gaussian kernel:
\begin{eqnarray}
k (\mathbf{X}_n, \mathbf{X}_m) &=& \exp \left( - \frac{1}{2 \sigma^2} \| \mathbf{X}_m - \mathbf{X}_n \|^2 \right),
\end{eqnarray}
where $\sigma$ is a length-scale parameter. 
The optimal values for the hyperparameters $\lambda$ and $\sigma$ need to be carefully chosen, which will be explained in the next subsection.

\section*{\sffamily \Large Cross-validation}

To obtain a good ML model with KRR, one should carefully determine the optimal values for the hyperparameters $\lambda$ and $\sigma$. 
In the present study, cross-validation schemes were used.\cite{Hansen-2013} 
In $S$-fold cross-validation (in this work, $S=10$), the data set $D = \{\mathbf{X}_n, F (\mathbf{X}_n) \}$ is randomly split into equally sized $S$ groups (or bins): $D_s$ with $s=1, ...,S$. 
One group is used as a test data set whereas the remaining $S-1$ groups are regarded as a training data set; as a consequence, the number of the test data and the number of the training data are $N_{\rm te}:=N/S$ and $N_{\rm tr}:=N (S-1) /S$, respectively.

For each data subset $D\setminus D_s$ consisting of $N_{\rm tr}$ samples (where $B\setminus A$ denotes the relative complement of A in B), we train a model using the KRR method and predict $F^{*(s)} (\mathbf{X}; \lambda, \sigma)$ that can depend on the hyperparameters $\lambda$ and $\sigma$. 
For each data subset $D_s$, the prediction error $\Delta^{(s)} (\lambda,\sigma)$ is estimated as the mean square error:
\begin{eqnarray}
\Delta^{(s)} (\lambda,\sigma) = \frac{1}{N_{\rm te}}
\sum_{l \, \in \, D_s}
\left[ F (\mathbf{X}_{l}) - F^{*(s)} (\mathbf{X}_{l}; \lambda, \sigma) \right]^2.
\label{delta_l}
\end{eqnarray}
The cross-validation error can be obtained by averaging $S$ different prediction errors. 
By minimizing the cross-validation error with respect to $\lambda$ and $\sigma$, one can find the optimal values of the hyperparameters, $\lambda^*$ and $\sigma^*$. 
Now that the final model is the one that gives the smallest $\Delta^{(s)} (\lambda^*, \sigma^*)$ among $S$ training data sets.

\section*{\sffamily \Large Representation of atomic configurations}

A number of descriptors have been developed to represent atomic environments: Coulomb matrix,\cite{Rupp-2012,Hansen-2013}, bispectrum\cite{Bartok-2010,Bartok-2013} and symmetry functions,\cite{Behler-2007,Behler-2011,Behler-2011a} to name but a few. 
Recently, an atomic fingerprint function suggested by Botu and Ramprasad has been shown to be a good descriptor in predicting atomic forces of solid systems\cite{Botu-2015a,Botu-2015b}; a similar descriptor has been independently proposed by Li, Kermode, and De Vita.\cite{Li-2015}
In the following, we consider a system made up of single atom species (an extension to multi-component systems has also been discussed in the literature\cite{Botu-2015a,Botu-2015b,Li-2015}). 
To efficiently represent the force acting on atom $i$ with the position $\left( x_i^u, y_i^u, z_i^u \right)$ at the configuration $u$, $\left(F_{x, i}^u, F_{y, i}^u, F_{z, i}^u\right)$, one may use an atom-centered fingerprint function for each Cartesian component\cite{Botu-2015a,Botu-2015b}:
\begin{eqnarray}
X_i^u \left(\eta \right) &=& \sum_{j \neq i} \frac{x_j^u - x_i^u}{r_{ij}^u} \exp \left[- \left(r_{ij}^u/\eta \right)^2 \right] f \left(r_{ij}^u \right), \ \ \ \ \\
Y_i^u \left(\eta \right) &=& \sum_{j \neq i} \frac{y_j^u - y_i^u}{r_{ij}^u} \exp \left[- \left(r_{ij}^u/\eta \right)^2 \right] f \left(r_{ij}^u \right), \ \ \ \ \\
Z_i^u \left(\eta \right) &=& \sum_{j \neq i} \frac{z_j^u - z_i^u}{r_{ij}^u} \exp \left[- \left(r_{ij}^u/\eta \right)^2 \right] f \left(r_{ij}^u \right), \ \ \ \ 
\end{eqnarray}
where the distance $r_{ij}^u$ is the Euclidian norm between atoms $i$ and $j$ at the configuration $u$, and $\eta$ determines the decay rate.
The function $f \left(r_{ij}^u \right)$ is a damping function that smoothly vanishes at a certain cutoff radius. In this work, $f \left(r_{ij}^u \right)$ is given by\cite{Behler-2007,Behler-2011,Behler-2011a,Botu-2015a}
\begin{eqnarray}
f \left( r_{ij}^u \right) &=& 0.5 \left[ \cos \left(\pi r_{ij}^u /R_{\rm c} \right) + 1 \right] 
\end{eqnarray}
for $r_{ij}^u \le R_{\rm c}$ and zero otherwise, where $R_{\rm c}$ is a cutoff radius. 
Different values for $R_{\rm c}$ will be investigated in the Results and Discussion section. 
In practice, the atomic fingerprint for each Cartesian component is given by a $K$-dimensional vector: $\mathbf{X}_i^u = (X_i^u (\eta_1) \cdots X_i^u (\eta_K))^{\rm T}$ for the $x$-component, with similar definitions for $\mathbf{Y}_i^u$ and $\mathbf{Z}_i^u$;
and a set of different values for $\eta$, $\{ \eta_{k} \} \, (k=1, ..., K)$, efficiently captures the local atomic configurations centered on a reference atom. 

When interpolating atomic forces from local atomic configurations, one needs to define the distance between two local environments, since the KRR method is based on the principle of similarity. 
To this end, one may use the Euclidean distance between two atomic fingerprint vectors,\cite{Botu-2015a,Botu-2015b} although other metrics for the distance can also be applied.\cite{Li-2015}
The distance between two local atomic configurations for the $x$-component may be defined by
\begin{eqnarray}
\left\| \mathbf{X}_i^u - \mathbf{X}_j^v \right\|
= \sqrt{\sum_{k=1}^K \left[ X_i^u (\eta_k) - X_j^v (\eta_k) \right]^2}, \label{distance_x}
\end{eqnarray}
with similar definitions for the $y$- and $z$-components. 
The distances among the atomic configurations are necessary for evaluating the kernel matrix between training data, and an interpolative prediction of each component of the atomic force can be obtained by a sum of weighted kernel functions, which can be computed using all the distances between a new atomic fingerprint vector and all the training ones.

\section*{\sffamily \Large Generating data and evaluating the force error}

To generate a variety of data sets for KRR, we performed DFT-based MD simulations on crystalline silicon at different temperatures: 300, 450, 600, 750, 900, 1200, 1500, and 1650\,K for a 64-atom system; 300, 900, and 1650\,K for a 512-sytem. 
The electronic structure calculations were carried out using a non-self-consistent tight-binding method, in which the total energy was evaluated by the Harris--Foulkes functional\cite{Harris-1985,Foulkes-1989,Miyazaki-2004} within the local-density approximation to the Kohn--Sham density functional theory.\cite{Kohn-1965}
We used a norm-conserving pseudopotential\cite{Troullier-1991} for Si to treat valence-core interactions and a single-$\zeta$ basis set with an energy grid cutoff of 108\,Hartree. 
Only the $\Gamma$ point was used to sample the Brillouin zone.

We used a cubic supercell of length $L=10.86 \, {\rm \AA}$ (the density $\rho = 2.33$\,g\,cm$^{-3}$) for a 64-atom system and a cubic supercell of length $L=21.72 \, {\rm \AA}$ for a 512-atom system, with periodic boundary conditions. 
We performed each simulation for 10 ps with a time step of 0.5 fs. 
To generate the canonical ensemble, we employed the Nos\'{e}--Hoover chain method,\cite{Martyna-1992,Martyna-1996} in which a chain of 5 thermostats with a thermostat frequency of 500\,cm$^{-1}$ was coupled to the ionic motions. 
The 15th-order Yoshida--Suzuki integrator was used to propagate the thermostat part of the time-reversible Liouville operator.\cite{Martyna-1996} 
The relative errors with respect to target temperatures were below 0.5\% for all the temperatures. 
All the simulations were performed using the C{\footnotesize ONQUEST} code.\cite{conquest,Bowler-2002,Bowler-2006}

To obtain an ML model with KRR, we created a training data set from the MD trajectory at 1650\,K because of its most expanded configuration space among all the simulations: we selected $N$ force data $(F_{x, i}^u, F_{y, i}^u, F_{z, i}^u)$ from the time region between 2,001 and 10,000 steps (i.e., 1--5\,ps), where the integers $i$ and $u$ were randomly chosen. 
Next, we created test data sets consisting of 10,000 atomic configurations taken from 10,001 to 20,000 steps (i.e., 5--10\,ps) at each temperature. 
After obtaining the optimal model using the KRR method together with the cross-validation scheme, we evaluated the prediction error for the atomic forces as the mean absolute error (MAE):
\begin{eqnarray}
\Delta F &=& \frac{1}{3N_{\rm a} N_{\rm av}} \sum_{i} \sum_{u}
\left[ \left| F_{x,i}^{u} -F_{x}^{*} (\mathbf{X}_i^u) \right| + \left| F_{y,i}^{u} -F_{y}^{*} (\mathbf{Y}_i^u) \right| + \left| F_{z,i}^{u} -F_{z}^{*} (\mathbf{Z}_i^u) \right| \right],
\end{eqnarray}
where the first sum runs over all the atoms in the supercell and the second sum all the atomic configurations during 5--10\,ps; the integer $N_{\rm a}$ is the number of atoms and the integer $N_{\rm av}$ the total number of the atomic configurations. 
The function $F_{x}^{*} (\mathbf{X}_i^u)$ is the predicted value for the $x$-component of the atomic force with the fingerprint vector $\mathbf{X}_i^u$ and similar definitions apply to $F_{y}^{*} (\mathbf{Y}_i^u)$ and $F_{z}^{*} (\mathbf{Z}_i^u)$.
The force error $\Delta F_{\rm av}$ was obtained by averaging over $32$ different choices for training data sets.


\section*{\sffamily \Large RESULTS AND DISCUSSION}

\section*{\sffamily \Large Atomic forces and atomic fingerprints}

ML is a purely data-driven, interpolative method; in other words, it is not guaranteed that extrapolation can predict properties as reliably as interpolation can do. 
For example, an ML model trained by a data set taken from the MD trajectory at 300\,K may not faithfully predict atomic forces at higher temperatures, since such a training data set does not contain highly distorted local atomic configurations or strong atomic forces caused by elevated thermal fluctuations. 
On the other hand, generating a number of QM data sets at different temperatures is computationally demanding. 
To circumvent such extrapolation issues as well as computational burden, we present a simple, intuitive prescription: 
to use a data set taken from the MD trajectory at a higher temperature, because such a data set is likely to be a physically relevant one, in which possible atomic displacements associated with normal modes as well as anharmonic effects are implicitly included.

Motivated by this picture, we investigated the histograms of the atomic forces in terms of temperature (Fig.~\ref{fig:distribution_f}a). 
We varied the temperature from 300 to 1650\,K (which is lower than the experimental melting point\cite{Kittel-2005}). 
For all the temperatures, the histograms showed Gaussian-like distributions centered on the origin; and the atomic forces at 1650\,K were most broadly distributed than those at lower temperatures, meaning that the upper and lower values for the atomic forces were bounded by those at the highest temperature. 
This can be quantitatively confirmed by the standard deviation $\delta$ of the atomic forces as a function of temperature (Fig.~\ref{fig:distribution_f}b). 
We also checked the atomic forces of a 512-atom system and found that the distributions for the larger system were very similar to those for a 64-atom system (not shown), ensuring the transferability to larger systems. 
For the sake of later discussions, it is convenient to define {\it force range} as $[-2.5 \delta, 2.5 \delta]$, in which about 99\% of entries lie,\cite{note_error} if the normal distribution is assumed. 
For instance, the force range at 300\,K can be calculated as $[-1.693, 1.693]$ in eV/\AA.

To validate the atomic fingerprints as an adequate descriptor for interpolation in terms of temperature, we also investigated the histograms of the atomic fingerprints at various temperatures (Fig.~\ref{fig:distribution_V}a). 
The distributions of the atomic fingerprints were qualitatively similar to those observed for the atomic forces; and as was the case for the atomic forces, the upper and lower values for the atomic fingerprints were bounded by those at 1650\,K. 
This was true for all the $\eta$ values examined, where the standard deviations of the atomic fingerprints for lower temperatures were bounded by the one at the highest temperature (Fig.~\ref{fig:distribution_V}b). 
Note that the standard deviation of the atomic fingerprints was more sensitive to smaller $\eta$ values than to larger ones. 
Our observations suggest a merit of using a training data set taken from the MD trajectory at an upper bound of the temperature range; in this study, we therefore trained our ML model using a data set generated at 1650\,K, the details of which will be described in the next subsection.

\section*{\sffamily \Large Training a model}

To accurately predict the atomic forces using the atomic fingerprints and KRR, we have to carefully choose two key parameters: the fingerprint complexity (a proper set of $\eta_{k}$) and the training data set size $N$. 
We started by addressing the fingerprint complexity with a fixed training size of $N=1000$. 
To our knowledge, detailed information about determining an optimal set of $\eta_{k}$ values has not been well reported. Here we used a set of equally spaced $\eta$ values within a given cutoff radius $R_{\rm c}$: $\eta_{k} = R_{\rm c} k / K \, (k=1, ..., K)$. 
This means that the task of finding a proper set of $\eta_{k}$ can be reduced to determining a fingerprint vector size $K$ within an appropriate cutoff radius.

Accordingly, we investigated the force error $\Delta F_{\rm av}$ by changing $R_{\rm c}$ from $2.72 \, {\rm \AA}$ up to the size of the supercell, $10.86 \, {\rm \AA}$. 
With a cutoff of $2.72 \, {\rm \AA}$, the force error was about $0.4 \, {\rm eV/\AA}$ or 5.3\% error with respect to the corresponding force range, resulting in the worst performance among all the cases (Fig.~\ref{fig:vec_dep_1650}). 
This is because the atomic fingerprint with such a short cutoff could not properly capture the essential information about the nearest atoms. 
With $R_{\rm c}$ larger than about $3 \, {\rm \AA}$ (which roughly corresponds to the first minimum in the radial distribution function; data not shown), the prediction errors were about $0.15 \, {\rm eV/\AA}$, or 2\% error. 
Obviously, we needed to increase the fingerprint vector size $K$ as we increased $R_{\rm c}$, in order to achieve similar performance (Fig.~\ref{fig:vec_dep_1650}). 
Using relatively large $R_{\rm c}$ did not improve the prediction accuracy (Figure~\ref{fig:vec_dep_1650}); for this reason, we chose $K = 10$ with a cutoff of $R_{\rm c} = 3.26 \, {\rm \AA}$ in the present study. 
While this value for the cutoff radius is shorter than the previously used value ($8 \, {\rm \AA}$),\cite{Botu-2015a,Caccin-2015} our ML model succeeded in predicting the QM forces with good performance. 
In general, the proper value for the cutoff radius may depend on materials or chemical species (e.g., long-range correlations may be more important for multi-component systems with polarization); nevertheless, our results may imply that accurate description of the nearest atoms plays a primal role in mapping atomic fingerprints to atomic forces of crystalline compounds.

Having determined an optimal set of $\eta_{k}$, we then plotted the force error as a function of the number of training data (Fig.~\ref{fig:num_dep_1650}). 
The prediction error $\Delta F_{\rm av}$ asymptotically decreased with increasing the size $N$ of the training data set.
Even with a training size of $N=200$, the force error was about 0.16\,eV/\AA, or 2.1\% error, indicating the efficiency of our ML model as well as our data-selection scheme. 
The error is substantially smaller than an estimated force error for the Stillinger--Weber potential at 1000\,K (about 0.5\,eV/\AA).\cite{Li-2015} 
In the case of $N=1000$, the force error at 1650\,K was about 0.15\,eV/\AA, or 2.0\% error with respect to the force range. 
We note that about 1000 configurations are sufficient to capture the essential information about local atomic environments for predicting atomic forces.\cite{Li-2015,Botu-2015b} 
This is also important because the computational cost required for the training phase in KRR scales cubically with respect to the number of training data.\cite{Hansen-2013,Botu-2015a} 
To balance computational effort with prediction accuracy, we trained an ML model using a training size of $N=1000$, with a fingerprint vector size of $K=10$ and $R_{\rm c} = 3.26 \, {\rm eV/\AA}$.

\section*{\sffamily \Large Model performance}

In this subsection, we address the performance and transferability of our ML model in predicting atomic forces at a wide range of temperatures. 
To this end, we applied the ML model (which was trained in the procedure described earlier) to a number of data sets taken from the canonical-ensemble MD trajectories at various temperatures. 
Not surprisingly, as temperature increased, so did the force range and the force error. The force error increased from 0.059 to 0.154\,eV/\AA \ as the temperature was changed from 300 to 1650\,K (Table 1). 
Here, we compare the force errors as a percentage of the corresponding force range.
The force errors were below 2\% at 300--1200\,K and remained about 2\% at 1500 and 1650\,K (Table 1). 
The results suggest the robustness of our ML model at a wide range of temperatures.

Our approach presented here is based on the premise that atomic forces should depend only on the local atomic environments, which indicates the transferability to larger system sizes.
To verify this view numerically, we also applied the same ML model to a 512-atom system at three different temperatures: 300, 900, and 1650\,K (Table 2). 
The force errors for a 512-atom system were quantitatively similar to those for a 64-atom system (see Tables~1 and 2), demonstrating that the prediction accuracy of atomic forces is independent of the global frame of reference. 
Our results agree with a recent ML study on QM properties of atoms in molecules.\cite{Rupp-2015}
Note that the computational effort of the ML evaluation of atomic forces scales linearly in system size.

Figure~\ref{fig:time_dep_force} demonstrates the performance and transferability of our ML model along MD trajectories for the two system sizes: 
(i) $T=1650$\,K and $N_{\rm a}=64$; (ii) $T=300$\,K and $N_{\rm a}=64$; (iii) $T=1650$\,K and $N_{\rm a}=512$; and (iv) $T=300$\,K and $N_{\rm a}=512$. In all the cases, the predictions of the atomic forces were excellent, showing the transferability with respect to temperature as well as system size. 
Our results indicate that useful and practical ML models could be trained by data sets taken from DFT-based MD simulations on smaller systems at a higher temperature and that ML models of this kind may be useful for performing MD simulations on large, realistic systems at various temperatures and for calculating their thermodynamic and kinetic properties.


\section*{\sffamily \Large SUMMARY}

By using the KRR method together with the atomic fingerprints, we trained an ML model on crystalline silicon system to directly predict the atomic forces in an interpolative manner. 
From a physical standpoint, we gave a simple, intuitive prescription to generate a versatile training data set for interpolation: the idea is that interpolation can be made by using merely a QM data set generated at a higher temperature, or an upper bound of the temperature range of interest. 
To verify this, we trained an ML model on a 64-atom system using a data set taken from the MD trajectory at 1650\,K and applied the ML model to predict the atomic forces in the temperature range from 300 to 1650\,K. 
The force errors between ML and QM evaluations were about 2\% or smaller, 
demonstrating the accuracy and robustness of our ML model. 
We also confirmed the applicability of our ML model to a larger system (a 512-atom system), showing that the prediction accuracy is independent of the global frame of reference. 
Our results suggest that, once the ML of QM forces at a higher temperature is conducted with a careful cross-validation scheme, interpolation of atomic forces can be made with adequate accuracy for various temperatures and system sizes. 
Our results imply that practical ML models could be trained by QM data sets obtained from MD simulations at a higher temperature and that ML models of this kind may be useful for performing MD simulations on large, realistic systems and for calculating their thermodynamic and kinetic properties.

\subsection*{\sffamily \large ACKNOWLEDGMENTS}

T. S. and R. T. thank Masato Sumita for helpful discussions. 
T. S. and T. M. were partly supported by JSPS KAKENHI project (Grants No. 26610120 and No. 26246021).
R. T. was partially supported by Nippon Sheet Glass Foundation for Materials Science and Engineering. 
The MD simulations in this study were performed on Numerical Materials Simulator at National Institute for Materials Science.
The calculations for machine learning were performed on the supercomputer at Supercomputer Center, Institute for Solid State Physics, the University of Tokyo. 
This work was supported by the ``Materials Research by Information Integration'' Initiative of the Support Program for Starting Up Innovation Hub, Japan Science and Technology Agency
and by  the WPI Initiative on Materials Nanoarchitectonics, Ministry of Education, Culture, Sports, Science and Technology of Japan.

\clearpage



\begin{thebibliography}{99}
\bibitem{note_error}
Instead of using the difference between the maximum and the minimum, which can be highly fluctuated across data sets, we used a similar definition for the property's range given by Rupp et. al.\cite{Rupp-2015a}
\end{thebibliography}


\clearpage

\begin{figure}
\caption{\label{fig:distribution_f}
(Left) Temperature dependence of the histograms of the atomic forces, $F_{x, i}^u$, $F_{y, i}^u$, and $F_{z, i}^u$, of a crystalline Si system consisting of 64 atoms. The histograms were normalized for comparison. 
(Right) Standard deviation $\delta$ of the atomic forces as a function of temperature.
}
\end{figure}

\begin{figure}
\caption{\label{fig:distribution_V}
(Left) Temperature dependence of the histograms of the atomic fingerprints, $X_{i}^u (\eta)$, $Y_{i}^u (\eta)$, and $Z_{i}^u (\eta)$, of a crystalline Si system consisting of 64 atoms (where $\eta=3.258 \, {\rm \AA}$). The histograms were normalized for comparison. 
(Right) $\eta$ dependence of the standard deviation of the atomic fingerprints in the temperature range of 300 to 1650 K.
}
\end{figure}

\begin{figure}
\caption{\label{fig:vec_dep_1650}
Force error $\Delta F_{\rm av}$ as a function of the fingerprint vector size, $K$, for various values for the cutoff radius $R_{\rm c}$. The number of training data is 1000.
Although we calculated the standard errors of these data,
the error bars are omitted for clarity since their ranges are smaller than the symbol size.
}
\end{figure}

\begin{figure}
\caption{\label{fig:num_dep_1650}
Force error $\Delta F_{\rm av}$ as a function of the number of training data $N$. A fingerprint vector size of $K=10$ and a cutoff of $R_{\rm c}=3.26 \, {\rm \AA}$ were used.
The error bars represent standard errors.
}
\end{figure}

\begin{figure}
\caption{\label{fig:time_dep_force}
Comparison of the QM and ML atomic forces as a function of time. Shown is the $x$-component of the force acting on a particular atom in crystalline silicon. (i) $T=1650$\,K and $N_{\rm a}=64$; (ii) $T=300$\,K and $N_{\rm a}=64$; (iii) $T=1650$\,K and $N_{\rm a}=512$; and (iv) $T=300$\,K and $N_{\rm a}=512$. For all the cases, we used the same ML model trained by a QM data set taken from the MD trajectory for a 64-atom system at 1650\,K.
}
\end{figure}



\clearpage

\vspace*{0.1in}   
\begin{center}
\includegraphics[scale=1.6]{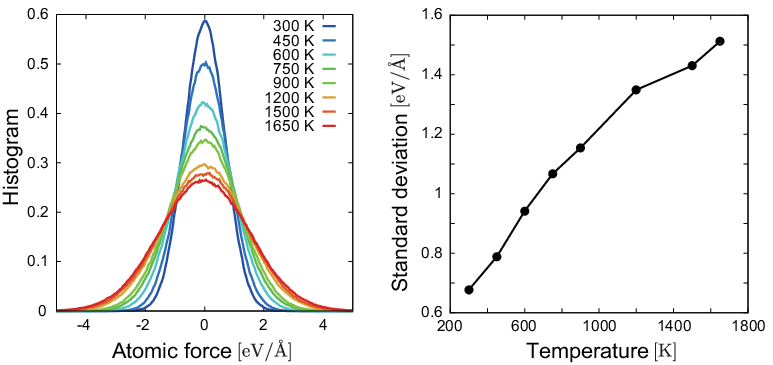} 
\end{center}
\vspace{0.25in}
\hspace*{3in}
{\Large
\begin{minipage}[t]{3in}
\baselineskip = .5\baselineskip
Figure 1 \\
Teppei Suzuki, Ryo Tamura, Tsuyoshi Miyazaki  \\
Int. J.\ Quant.\ Chem.
\end{minipage}
}

\vspace*{0.1in}   
\begin{center}
\includegraphics[scale=1.6]{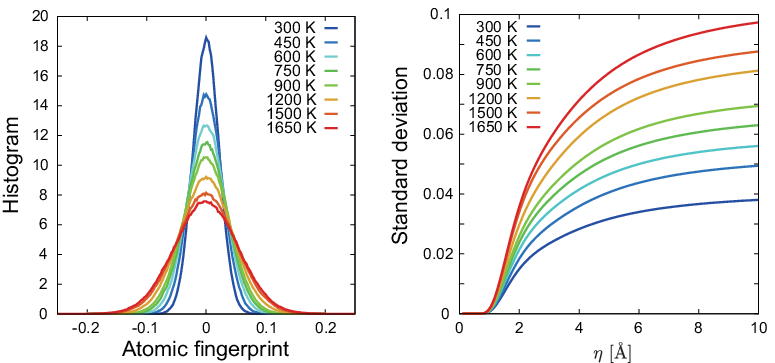} 
\end{center}
\vspace{0.25in}
\hspace*{3in}
{\Large
\begin{minipage}[t]{3in}
\baselineskip = .5\baselineskip
Figure 2 \\
Teppei Suzuki, Ryo Tamura, Tsuyoshi Miyazaki  \\
Int. J.\ Quant.\ Chem.
\end{minipage}
}

\vspace*{0.1in}   
\begin{center}
\includegraphics[scale=1.4]{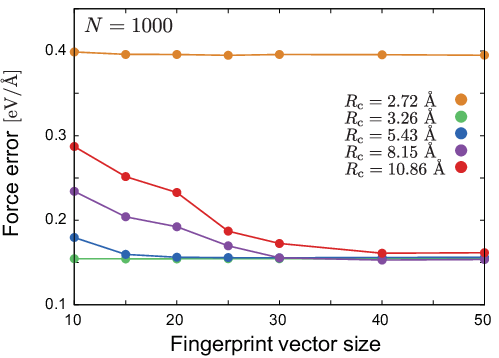} 
\end{center}
\vspace{0.25in}
\hspace*{3in}
{\Large
\begin{minipage}[t]{3in}
\baselineskip = .5\baselineskip
Figure 3 \\
Teppei Suzuki, Ryo Tamura, Tsuyoshi Miyazaki  \\
Int. J.\ Quant.\ Chem.
\end{minipage}
}

\vspace*{0.1in}   
\begin{center}
\includegraphics[scale=1.4]{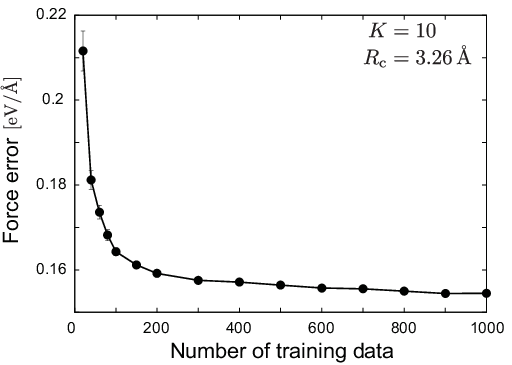} 
\end{center}
\vspace{0.25in}
\hspace*{3in}
{\Large
\begin{minipage}[t]{3in}
\baselineskip = .5\baselineskip
Figure 4 \\
Teppei Suzuki, Ryo Tamura, Tsuyoshi Miyazaki  \\
Int. J.\ Quant.\ Chem.
\end{minipage}
}

\vspace*{0.1in}   
\begin{center}
\includegraphics[scale=1.0]{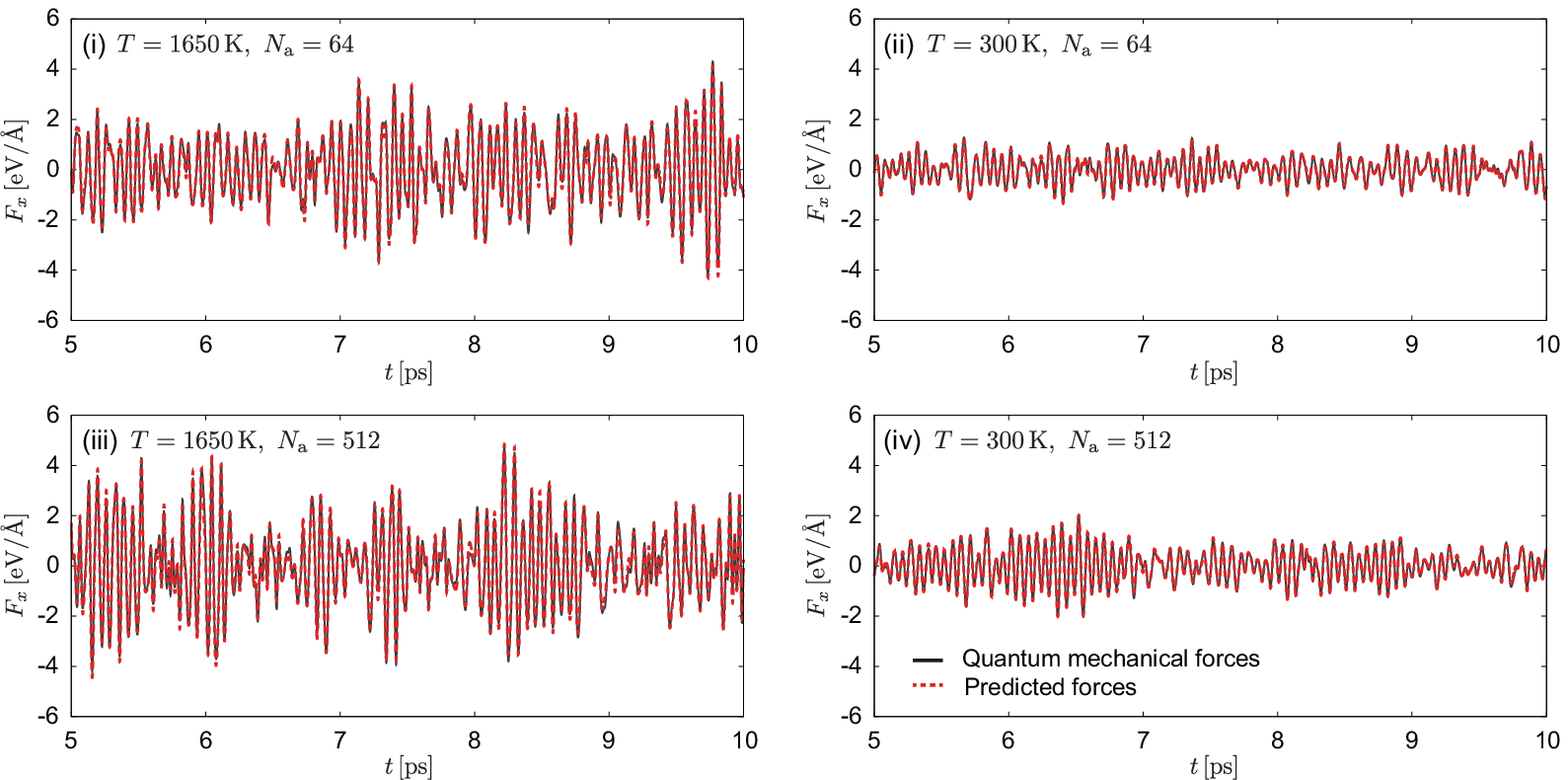} 
\end{center}
\vspace{0.25in}
\hspace*{3in}
{\Large
\begin{minipage}[t]{3in}
\baselineskip = .5\baselineskip
Figure 5 \\
Teppei Suzuki, Ryo Tamura, Tsuyoshi Miyazaki  \\
Int. J.\ Quant.\ Chem.
\end{minipage}
}

\clearpage

\begin{table}
\begin{tabular}{|c|c|c|c|}
\hline
\textbf{Temperature [K]} & \textbf{Range [eV/\AA]} & \textbf{Error [eV/\AA]} & \textbf{Ratio [\%]} \\ 
\hline
 300&       $\pm1.693$&     0.059&       1.72 \\ \hline
 450&      $\pm 1.971$ &     0.072&       1.83 \\ \hline
 600&       $\pm 2.354$ &     0.081&       1.72 \\ \hline
 750&       $\pm 2.667$ &     0.090&       1.69 \\ \hline
 900&       $\pm 2.885$ &     0.102&       1.77 \\ \hline
1200&       $\pm 3.373$&     0.116&       1.73 \\ \hline
1500&       $\pm 3.575$&     0.147&       2.05 \\ \hline
1650&       $\pm 3.780$&     0.154&       2.04 \\
\hline
\end{tabular}
\caption{\label{tab:64}
Force ranges and the force errors $\Delta F_{\rm av}$ for a 64-atom system. Also presented is the ratio of the force error with respect to the corresponding force range. Each force range was defined by $\pm 2.5 \delta$, with $\delta$ being the standard deviation of the atomic forces at each temperature. See also Figure~\ref{fig:distribution_f}b.
}
\end{table}

\begin{table}
\begin{tabular}{|c|c|c|c|}
\hline
\textbf{Temperature [K]} & \textbf{Range [eV/\AA]} & \textbf{Error [eV/\AA]} & \textbf{Ratio [\%]} \\ 
\hline
 300&       $\pm 1.657$ &     0.057&       1.72 \\ \hline
 900&      $\pm 2.816$ &     0.100&       1.78 \\ \hline
1650&       $\pm 3.760$ &     0.143&       1.90 \\
\hline
\end{tabular}
\caption{\label{tab:512}
Force ranges and the force errors $\Delta F_{\rm av}$ for a 512-atom system. Also presented is the ratio of the force error with respect to the corresponding force range. Each force range was defined by $\pm 2.5 \delta$, with $\delta$ being the standard deviation of the atomic forces at each temperature.
}
\end{table}

\end{document}